\documentclass[%
twocolumn,%
oneside,%
floats,%
aps,%
prd,%
nobibnotes,%
nofootinbib,%
amsmath,%
amssymb,%
amsfonts,%
amscd,%
showpacs,%
eqsecnum%
]{revtex4-1}

\usepackage[utf8]{inputenc}
\usepackage{graphicx,array,dcolumn}
\usepackage[paperwidth=210mm,paperheight=297mm,centering,hmargin=1.8cm,vmargin=2.5cm]{geometry}

\newcommand{\be}{\begin{equation}}
\newcommand{\ee}{\end{equation}}
\def\larghezza{.36\textwidth}

\def\b{\textbf}
\def\i{\textit}
\def\mb{\mathbf}
\def\mc{\mathcal}
\def\be{\begin{equation}}
\def\ee{\end{equation}}
\def\mpl{m_{Pl}}
\def\mn{{\mu\nu}}
\def\D{\mc D}
\def\-g{\sqrt{-g}}

\renewcommand\rho{\varrho}
\renewcommand\tilde{\widetilde}
\newcommand\rif[1]{(\ref{#1})}
\def\so{\Rightarrow}
\newcommand{\rhoc}{10^{-29}\text{ g cm}^{-3}}
\def\R{R}
\def\T{T}

\begin{document}
\author{Lorenzo Reverberi}
\email{reverberi@fe.infn.it}
\title{Curvature Singularities from Gravitational Contraction in $f(\R)$ Gravity}
\affiliation{Dipartimento di Fisica e Scienze della Terra, Universit\`a degli Studi di Ferrara\\
Polo Scientifico e Tecnologico - Edificio C, Via Saragat 1, 44122 Ferrara, Italy}

\affiliation{Istituto Nazionale di Fisica Nucleare (INFN), Sezione di Ferrara\\
Polo Scientifico e Tecnologico - Edificio C, Via Saragat 1, 44122 Ferrara, Italy}

\begin{abstract}
The discovery of the accelerated expansion of the Universe has had a vast resonance on a number of physical disciplines. In recent years several viable modified gravity models have been proposed, which naturally lead to a late-time de Sitter stage while basically reducing to General Relativity in the early Universe. We consider a contracting cloud of pressureless dust, having arbitrary mass and initial density, and study some aspects of these modified gravity models. We show how the increasing energy/mass density may lead to a curvature singularity and discuss the typical timescales for its development. \\
\end{abstract}

\maketitle

%


\section{Introduction}
The physical mechanism behind the present accelerated cosmological expansion \cite{Nobel_2011} is still unknown. A pure cosmological constant term is quite natural both in General Relativity and in Quantum Field Theory, but then there remains the challenge of explaining the present value of $\Lambda$, in particular its smallness and the fact that $\Omega_m\sim \Omega_\Lambda$ (coincidence problem).

More complicated but popular scenarios involve ``quintessence'' models \cite{Fujii_82_and_Ratra_Peebles_88_and_Caldwell_98_and_Wetterich_98}, in which a scalar field coupled to gravity is responsible for the acceleration, or modified gravity models, in which the acceleration is due to modifications of the Einstein-Hilbert action and therefore of the Einstein field equations. In the simplest case, the gravitational lagrangian is a non-linear function of the scalar curvature $\R$ alone:\footnote{We use natural units $c=\hbar=k=1$, and the Planck mass is defined as $\mpl^2=G_N^{-1}$. The metric has signature $(+\,-\,-\,-)$, and we use the conventions $\Gamma^\alpha_\mn = \frac{1}{2}g^{\alpha\beta}(\partial_\mu g_{\beta\nu}+\cdots)$, $\R^\alpha_{\,\,\mu\beta\nu}=\partial_\beta \Gamma^\alpha_\mn+\cdots $, $\R_\mn = \R^\alpha_{\,\,\mu\alpha\nu}$, $\R=\R^\mu_\mu$. With these conventions, $\R<0$ for a matter-dominated Universe.}
\be\label{eq:grav_action}
\begin{aligned}
A_{grav}&=-\frac{\mpl^2}{16\pi}\int d^4x\,\-g\,f(\R)\\
&\equiv -\frac{\mpl^2}{16\pi}\int d^4x\,\-g\,\left[\R+F(\R)\right]\,.
\end{aligned}
\ee
Following the seminal paper~\cite{Starobinsky_1980}, where it was found that terms $\sim \R^2$ appear naturally from one-loop corrections to the matter energy-momentum tensor, these models were initially discussed in the ultraviolet regime~\cite{ultraviolet}. In the first infrared-modified $f(\R)$ models proposed~\cite{Capozziello_2003_and_Carroll_Duvvuri_2004}, where $F\sim 1/\R$, the negative power of $\R$ determines the predominance of such term at late times, resulting in an accelerated Universe expansion.

Beside doubts about their viability and relevance~\cite{Chiba_2003}, these models were soon discovered to suffer from strong instabilities in the presence of gravitating bodies~\cite{DolgKaw}; see also~\cite{Faraoni_2006}.

The constraints for the cosmological viability of $f(\R)$ models were later thoroughly investigated \cite{Amendola_2007_and_Sawicki_2007}, and recently a few models have been proposed which evade all such tests, therefore seeming to be good candidates for a gravitational theory of Dark Energy~\cite{Appleby_Battye_2007, Hu_Sawicki_2007, Starobinsky_2007} (see also below, Eq.~\ref{eq:models}).

Testing modified gravity theories in astronomical/astrophysical systems is paramount to constrain and possibly rule out models, and in general to improve our knowledge of the subject. Studies of the stability of spherically symmetric solutions have indicated the possibility of an infinite-$\R$ singularity developing inside relativistic, dense stars~\cite{Frolov_2008_etc}. Important steps forward in our understanding of static, spherically or axisymmetric astrophysical objects in $f(\R)$ gravity have recently been made (see e.g.\cite{Babichev_Langlois_and_Capozziello_De_Laurentis_2012}), and seem to point towards the existence of a rather general instability/singularity problem in these theories. Indeed, it has been shown that analogous problems occur in many different extended theories of gravity, not only $f(\R)$~\cite{Seifert_2007}.

Furthermore, similar results are obtained in the case of a less dense but contracting object~\cite{Arb_Dolgov, Bamba_Nojiri_Odintsov_2011}. In this case the singularity is not triggered by the large mass/energy density, but rather from its increase with time. One can write the trace of the modified Einstein equations as an oscillator equation for the additional gravitational scalar degree of freedom, which is sometimes dubbed \i{scalaron} and which we will denote with $\xi$ (see below), and it is easy to see that $\R$ oscillates around the GR solution $\R+\T=0$. The frequency and the amplitude of such oscillations usually grow along with the increasing density, and may eventually lead to a singularity. The key point is that $\xi$ moves in a matter- and therefore time-dependent potential, in which the ``energy'' corresponding to the point $\xi=\xi(|\R|\to\infty)$ may be finite, rendering this singular point, in principle, accessible by the field. This mechanism is strictly related to that responsible for the past cosmological singularities examined e.g. in~\cite{
Appleby_Battye_2008, Appleby_Battye_Starobinsky_2010}. 

Singularity issues in infrared-modified $f(\R)$ theories could in principle be solved by the introduction of ultraviolet corrections, as investigated e.g. in~\cite{Arb_Dolgov,Bamba_Nojiri_Odintsov_2011}. Moreover, oscillations lead to gravitational particle production, and a large frequency/amplitude of the oscillating curvature could lead to a noticeable emission of cosmic rays~\cite{Arb_Dolg_Rev_2012}; in principle, this could severely affect the total cosmic ray flux, distort the power spectrum, and even serve as a possible mechanism to avoid the GZK cutoff~\cite{Arb_Dolg_Rev_preparation}.

In this paper, we focus again on the $f(\R)$ gravity models~\cite{Hu_Sawicki_2007,Starobinsky_2007} during the contraction of a nearly-homogeneous cloud of pressureless dust. Using a simplified approach, namely assuming spherical symmetry, homogeneity and low gravity, we work out simple expressions for the evolution of $\xi$ and hence $\R$, and in particular of the amplitude and frequency of their oscillations. We confirm the existence of a finite-time, future singularity, whose appearance depends on the duration of the contraction and on both model- and physical parameters, and derive estimates for the typical timescales for this process. 

Once we derive general results, we will apply them to two very similar models recently proposed and cited above:
\begin{subequations}\label{eq:models}
\begin{align}
 &F_{HS}(\R) = -\cfrac{\lambda \R_c}{1+(\R/\R_c)^{-2n}}\,,\quad\text{\cite{Hu_Sawicki_2007}}\\
 &F_S(\R) = \lambda \R_c\left[\left(1+\dfrac{\R^2}{\R_c^2}\right)^{-n}-1\right]\,.\quad \text{\cite{Starobinsky_2007}}
\end{align}
\end{subequations}
The subscripts stand for, respectively, Hu-Sawicki (HS) and Starobinsky (S). For both models, if $\lambda$ is of order unity $\R_c$ is of the order of the present cosmological constant (for details we refer the reader to the specific articles), which is much smaller than the typical values of $\R$ and $\T$ in astrophysical systems, such as pre-stellar, pre-galactic, and molecular clouds. Hence, in many cases we will take the limit $|\R_c/\T|\sim \R_c/\R \ll 1$ before presenting the final results.

For simplicity, we assume that the contraction of the system is stationary, i.e. the mass density grows linearly with time, on a typical timescale $t_{contr}$:
\be\label{eq:T_evol}
\T(t) = \T_0\left(1+t/t_{contr}\right)\,.
\ee
We must stress that this evolution law should not be regarded as accurate from a physical standpoint; the difficult task of computing the full dynamics of contraction of a self-gravitating system goes way beyond the scope of this paper (see e.g.~\cite{Cembranos_2012} and references therein).

Nevertheless, unless the contraction follows a radically different behaviour, and especially until ${t\sim t_{contr}}$, results obtained with this form should be qualitatively correct. We will find that a faster contraction contributes positively to the formation of a singularity, so we expect that contraction laws $\T\sim t^\gamma$ with $\gamma>1$ will lead to singularities even more effectively than what appears from our results. On the other hand, a slower contraction could help delaying ($\gamma<1$) or even avoiding, if the contraction would stop at some moment, the singularity. 

In this paper we will use the following dimensionless parameters characterising the physical properties of the system under scrutiny:
\be\label{eq:definitions}
\begin{aligned}
 \R_{29} &\equiv \frac{\mpl^2}{8\pi}\,\frac{(-\R_c)}{\rhoc}\,,\\
\rho_{29} &\equiv \frac{\rho_0}{\rhoc}\,,\\
t_{10} &\equiv \frac{t_{contr}}{10^{10}\text{ years}}\,.
\end{aligned}
\ee

\section{Curvature Evolution in Contracting Systems}\label{sec:CURV_EVOL}
\subsection{Field Equations}\label{sec:field_equations}
From Eq.~\rif{eq:grav_action}, one obtains the field equations
\be\label{eq:field_equations}
\begin{aligned}
f'(\R)&\R_\mn-\frac{1}{2}f(\R)\,g_\mn\,+\\
&+\left(g_\mn\D^2-\D_\mu\D_\nu\right)f'(\R)=\T_\mn\,.
\end{aligned}
\ee
Here, $\D$ denotes covariant derivative, a prime denotes derivative with respect to $\R$, and
\[
 \T_\mn \equiv \frac{8\pi}{\mpl^2}\,\frac{2}{\-g}\frac{\delta\left(\-g\,\mc L_m\right)}{\delta g^{\mn}}\,,
\]
where $\mc L_m$ is the matter lagrangian density. The corresponding trace equation reads
\be\label{eq:trace}
3\D^2F'+\R F'-2F-(\R+\T)=0\,.
\ee
We consider a nearly homogeneous and spherically symmetric cloud of pressureless dust, hence
\be
\T\simeq \frac{8\pi}{\mpl^2}\,\rho_m\,.
\ee and $\rho_m$ is the mass/energy density of the cloud. The homogeneity of the could allows us to neglect spatial derivatives, as intuitively clear and explicitly proved in~\cite{Arb_Dolgov}; assuming also low-gravity, the D'Alambertian operator can be replaced by the second derivative in the time coordinate: ${\D^2\to\partial_t^2}$. Of course, a more careful investigation of the problem should take into account both time and spatial derivatives; this could be a subject for further research. Also notice that these and following arguments can also be applied to cosmology, if we consider the evolution of the Universe, backwards in time, during the matter-dominated epoch~\cite{Appleby_Battye_2008, Appleby_Battye_Starobinsky_2010}. The approximation of low-gravity and the formation of curvature singularities may seem utterly incompatible, but this is not true: in fact, ${R\sim \partial^2 g_\mn}$ may diverge even if $g_\mn$ is very close to $\eta_\mn$ (Minkowski). Details about this statement and about the dust assumption (${p/\rho\ll 1}$) can be found in Sec.~\ref{sec:approx}.

As mentioned in the Introduction, in a typical astrophysical situation we have
\be\label{eq:large_density_limit}
\frac{\T}{\R_c}\sim\frac{\R}{\R_c} \gg 1\,,
\ee
so that both models reduce to
\be\label{eq:model_approx}
F_{HS}\simeq F_S\simeq-\lambda \R_c\left[1-\left(\frac{\R_c}{\R}\right)^{2n}\right]\,.
\ee
In this limit, these models are basically equivalent to adding a $\Lambda$-term, since $F$ is almost constant; in fact, we clearly have
\be\label{eq:Lambda}
\R+F\approx \R+2\Lambda\,,\qquad\Lambda = -\frac{\lambda \R_c}{2}\,.
\ee
See also~\cite{Hu_Sawicki_2007,Starobinsky_2007} for further details. Therefore, ${\R+\T+2F=0}$ is practically equivalent to GR with the addition of a cosmological constant $\Lambda$, that is the usual $\Lambda$CDM model\footnote{Of course, these $f(\R)$ models do \i{not} explain the nature of Dark Matter.}; in what follows, when referring to ``GR'' for brevity, we will mean precisely this. We also have ${F'\sim \lambda (\R_c/\R)^{2n+1}}$, thus
\be\label{eq:terms_com}
|\R| \gg |F| \gg |F'\R|\,.
\ee
Under these assumptions, Eq.~\rif{eq:trace} becomes
\be\label{eq:trace_approx}
3\partial_t^2F'-(\R+\T+2F)=0\,.
\ee
Since $F\sim\Lambda\ll T$, the only contribution of this effective cosmological constant is to offset the GR solution from $\R+\T=0$ to $\R+\T+2F=0$, which is a very small (and almost constant in absolute value) correction of order $|\R_c/\T|$. One should be careful, because it may appear from exact numerical results that $F$ is of the order of $\R+\T$ or even larger\footnote{But not of the order of $\R$ or $\T$ \i{individually}!}; nevertheless, its effect in~\rif{eq:trace_approx} is completely trivial, unlike the dynamical term $\ddot F'$, because even large variations of $\R$, especially when $|\R|$ increases, result in extremely small variations of $F$ (see equation~\ref{eq:model_approx}). From now on, we will include these corrections using
\be\label{eq:T_correction}
\tilde\T \equiv\T+2F \approx\T+4\Lambda\,,
\ee
However, we will still have to consider $\T$ alone because it is the quantity directly related to the \i{physical} energy/matter density at a given point and a given instant of time.\\
Defining the new scalar field
\be\label{eq:xi_def}
\xi\equiv -3F'\,,
\ee
which in the cases considered is approximately
\be\label{eq:xi_models_explicit}
\xi_{HS,\,S}\simeq 6n\lambda\left(\frac{\R_c}{\R}\right)^{2n+1}\,,
\ee
we rewrite~\rif{eq:trace_approx} as an oscillator equation:
\be\label{eq:trace_xi_KG}
\ddot\xi+\R+ \tilde\T=0\quad\Leftrightarrow \quad \ddot\xi+\frac{\partial U}{\partial\xi}=0\,.
\ee
Again, we stress the underlying assumption that ${\tilde\T=\tilde\T(t)}$. We are testing the behaviour of curvature with a simple, smooth \i{external} energy density evolution, which is arbitrarily chosen. As we have already mentioned, a more complete analysis could be subject of stimulating further research.

Usually, it is not possible to invert \rif{eq:xi_def} to obtain ${\R=\R(\xi)}$ and thus a simple form for $U(\xi)$, expect perhaps in some limit, but it is rather clear that solutions will oscillate around the solution ${\R+\tilde\T=0}$, with frequency roughly given by
\be
\omega_\xi^2\simeq \left.\frac{\partial^2 U}{\partial\xi^2}\right|_{\R+\tilde\T=0}\simeq \left.\frac{1}{\partial\xi/\partial \R}\right|_{\R+\tilde\T=0}\,.
\ee
If ${\omega^2 <0}$, one expects instabilities, and this is exactly the kind of instability of refs.~\cite{DolgKaw, Faraoni_2006}. One can immediately see that for the two models considered we have
\be\label{eq:omega_models}
\omega^2\simeq -\frac{\R_c}{6n\lambda(2n+1)}\left(-\frac{\tilde\T}{\R_c}\right)^{2n+2}>0\,,
\ee
so there is no instability problem. We remind the reader that with our sign conventions ${\R_c<0}$, ${\tilde\T>0}$.\\
Even with ${\omega^2>0}$, however, we will show that if the model fulfils some requirements, then curvature \i{singularities} can develop. In particular, we need that
\begin{itemize}
\item  there exist a certain value $\xi_{sing}$ corresponding to ${|\R|\to\infty}$,
\item the potential be finite in $\xi_{sing}$, i.e. ${U(\xi_{sing})<\infty}$.
\end{itemize}

\subsection{Energy Conservation and the Scalaron Potential}
If the previous requirements are met, then in general it is possible that $\xi$ reach $\xi_{sing}$ and hence ${|\R|\to\infty}$. We can see this, for instance, from the ``energy'' conservation equation associated to \rif{eq:trace_xi_KG}, that is
\be\label{eq:energy_conservation}
\frac{1}{2}\,\dot\xi^2+U(\xi,t)-\int^t dt'\,\frac{\partial \tilde\T}{\partial t'}\,\xi(t')=\text{ const}\,,
\ee
where
\be\label{eq:potential}
U(\xi,t) = \tilde\T(t)\,\xi + \int^\xi \R(\xi')d\xi'\,.
\ee
The last term in the l.h.s. of~\rif{eq:energy_conservation} is due to the explicit time-dependence of $\tilde\T$, and if ${\partial \tilde\T/\partial t>0}$, as is the case in contracting systems even without specifically assuming\footnote{Assuming \rif{eq:T_evol}, ${\partial\tilde\T/\partial t= \T_0/t_{contr}\neq \tilde\T_0/t_{contr}}$ (see~Eqs.~\ref{eq:T_evol} and~\ref{eq:T_correction}), so~\rif{eq:energy_conservation} is further simplified, with the last term simply being proportional to ${\int^t dt'\,\xi}$.}~\rif{eq:T_evol}, it will produce an increase in the ``canonical'' energy (kinetic + potential).

Note that this is true for $\xi>0$, whereas for $\xi<0$ it would give the opposite behaviour. However, it has been shown that the condition $F'<0$, corresponding to $\xi>0$, is crucial for the correct behaviour of modified gravity models at (relatively) low curvatures \cite{Appleby_Battye_Starobinsky_2010}.

\begin{figure}[t]
\includegraphics[width=.35\textwidth]{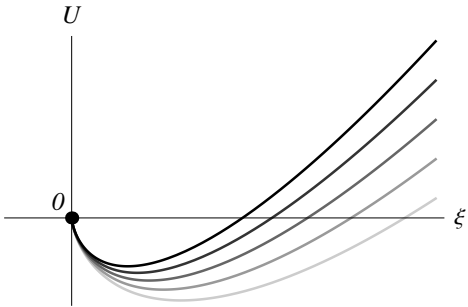}
\caption{Qualitative shape of potentials for models \rif{eq:models}, assuming ${\R/\R_c\gg 1}$. For both models ${\xi=\xi_{sing}=0}$ (the black dot) corresponds to the singular point ${|\R|\to\infty}$. The typical time evolution is also shown, from light grey (earlier times) to black (later times).}
\label{fig:potentials}
\end{figure}

As we have previously mentioned, usually it not possible to invert the relation ${\xi=\xi(\R)}$ in order to obtain ${\R=\R(\xi)}$ and solve the integral in \rif{eq:potential} exactly. Nonetheless, for the two models considered and in the limit ${\R/\R_c\gg 1}$ this procedure is possible; apart from an additive constant, which we can put to zero, the approximate potentials are equal and read
\be\label{eq:potentials_approx}
U\simeq \tilde\T(t)\,\xi + 3\lambda \R_c(2n+1)\left(\frac{\xi}{6n\lambda}\right)^\frac{2n}{2n+1}\,.
\ee
The shape of this potential is shown in Fig.~\ref{fig:potentials}. The bottom of the potential, as expected from \rif{eq:trace_xi_KG}, is in ${\R+\tilde\T=0}$. Moreover, ${\xi_{sing}=0}$ and ${U(\xi_{sing})=}$const.

\section{Adiabatic Region}\label{sec:adiabatic}
For simplicity, let us initially assume that the oscillations of $\xi$ in its potential are ``adiabatic'', in the sense that at each oscillation $\xi$ moves between two values
\be\label{eq:xi_min_xi_max}
\xi_{min}(t)\,,\quad\xi_{max}(t)\,,                                                                                                                                                                                                                                                                                                                           \ee
at roughly the same ``height'', that is with
\[
U(\xi_{min})=U(\xi_{max})\,.
\]
We should stress that $\xi_{min}$ and $\xi_{max}$ are considered to be slowly varying, so that it makes sense to compare these two values even though they do not correspond to the same instant of time, but are rather evaluated at different times with a $\delta t$ of order $\omega^{-1}$. 

The validity of this approximation can be understood as follows: the potential is roughly of the order of $\tilde\T\,\xi$, whereas the variation of the integral term in \rif{eq:energy_conservation} in one oscillation is of order $\dot{\tilde\T}\,\xi/\omega$. If $\omega$ is much larger than the inverse contraction time, that is
\[
\omega\gg \frac{\dot{\tilde\T}}{\tilde\T}\,,
\]
which is the case for the models considered below (see Eq.~\ref{eq:models}) provided that the contraction is sufficiently slow, then the integral term can be considered approximately constant over a large number of oscillations.\\
Assuming~\rif{eq:T_evol}, this basically results in the condition
\be\label{eq:fast-roll-condition}
\frac{\omega\,t_{contr}}{2\pi} \gg 1\,,
\ee
where the factor $2\pi$ only indicates that the period of the oscillations of $\xi$ is $2\pi\omega^{-1}$, not $\omega^{-1}$. For the models under investigation, this gives roughly
\be\label{eq:fast_roll_inequality}
\frac{\rho_{29}^{2n+2}\,t_{10}^2}{n\lambda(2n+1)\R_{29}^{2n+1}} \gg 142\,.
\ee
Later, we will relax this assumption and work in the opposite regime, where $\dot{\tilde \T}/\tilde\T \gtrsim \omega$.

Let us expand $\xi$ around the ``average'' value
\be\label{eq:xi_a_def}
\xi_a(t)\equiv \xi(\R=-\tilde\T)\,,
\ee
which corresponds to the value of $\xi$ if the behaviour of the system were described by the usual GR solution $\R+\tilde\T=0$.
We could be misled to infer from~\rif{eq:trace_approx} that with this definition $\xi_a$ must exactly satisfy
\be\label{eq:xi_a_wrong}
\ddot\xi_a=0\,,
\ee
so that $\xi_a \sim t$. This is not true, because near the GR solution we can no longer neglect sub-leading terms in \rif{eq:trace} and hence use \rif{eq:trace_approx}. In some sense, \rif{eq:xi_a_wrong} remains true provided that we interpret it as the statement
\be
\left|\ddot\xi_a\right| \ll |\R|,\,\tilde\T\,.
\ee 
Ultimately, $\xi_a$ is the reference point for $\xi$ because it corresponds to the bottom of its potential (see Eq.~\ref{eq:trace_xi_KG}). Nonetheless $\xi_a$ is not the solution of \rif{eq:trace_xi_KG}, but merely a test function helping us quantifying how the behaviour of $\R$ in $f(\R)$ gravity differs from that of GR. After all, $\xi$ in GR is identically zero.

Thus we write
\begin{subequations}\label{eq:xi_expansion}
\begin{align}
\xi(t) &= \xi_a(t)+\xi_1(t) \label{eq:xi_expansion_xi1}\\
&\equiv \xi_a(t) + \alpha(t)\sin \Phi(t)\,, \label{eq:xi_expansion_alpha}
\end{align}
where
\be
\Phi(t) \simeq \int^tdt'\,\omega\,.
\ee
\end{subequations}
The function $\alpha$ is also assumed to be relatively slowly-varying, that is
\be
\frac{\dot\alpha}{\alpha} \ll \omega\,.
\ee
In terms of the  quantities of~\rif{eq:xi_min_xi_max}, we have
\be\label{eq:xi_a_+-_alpha}
\xi_{min}\simeq \xi_a - \alpha\,,\quad \xi_{max} \simeq \xi_a+\alpha\,.
\ee

\subsection{Harmonic Regime}\label{sec:linearised}
We initially assume that the amplitude of oscillations is small enough that the potential can be approximated by a harmonic potential:
\be\label{eq:potential_approx_harmonic}
U(\xi,t) \simeq U_0(t)+\frac{1}{2}\,\omega^2(\xi-\xi_a)^2\,,
\ee
where $\omega$ was defined in~\rif{eq:omega_models} and as we can see from equations~\rif{eq:xi_models_explicit} and~\rif{eq:potentials_approx}
\be\label{eq:U0}
U_0(t) \equiv U\left(\xi_a(t)\right) = 3\lambda \R_c\left(-\frac{\R_c}{\tilde\T(t)}\right)^{2n}\,.
\ee
This is equivalent to considering the first-order approximation in $\xi_1$ (defined in Eq.~\ref{eq:xi_expansion_xi1}).  Equation~\rif{eq:trace_xi_KG} then reads
\be\label{eq:ddot_xi_1}
\ddot\xi_1+\omega^2\xi_1\simeq -\ddot\xi_a\,.
\ee
Using the expansion \rif{eq:xi_expansion_alpha} and neglecting $\ddot\xi_a$ and $\ddot\alpha$ yields
\be\label{eq:alpha_general_sol}
\frac{\dot\omega}{\omega}\simeq -2\frac{\dot\alpha}{\alpha}\quad\so\quad \alpha(t)\simeq \alpha_0\sqrt\frac{\omega_0}{\omega(t)}\,.
\ee
As long as the approximations hold, this can be considered a rather general result, and the specific $F(\R)$ model will determine the behaviour of the oscillations. The value $\alpha_0$ in Eq.~\rif{eq:alpha_general_sol} is strictly related to the initial conditions, that is to the initial displacement from the GR behaviour. We will fix the initial values of $\R$ and $\dot \R$, and from those derive the initial values of $\xi$ and $\dot\xi$. Thus, $\alpha_0$ can be calculated differentiating Eq.~\rif{eq:xi_expansion_alpha}, yielding
\be
\dot\xi_0(\R_0,\dot \R_0) \simeq \dot\xi_{a,0}+\alpha_0\omega_0\,\,\so\,\,\alpha_0\simeq \frac{\dot\xi_0-\dot\xi_{a,0}}{\omega_0}\,.
\ee
This corresponds to the explicit solution
\be\label{eq:sol_alpha_explicit}
\alpha(t) \simeq \left(\dot\xi_0-\dot\xi_{a,0}\right)\left[\omega_0\,\omega(t)\right]^{-1/2}\,.
\ee
Please note that, apparently, we have not made use of the assumption $U(\xi_{sing})<\infty$ considered before. Although not necessary to perform calculations, this condition is needed to ensure that the expansion \rif{eq:xi_expansion} be reliable. In fact, oscillations are harmonic only if the potential is nearly quadratic; this assumption is usually quite reasonable, especially near the bottom of the potential, but loses validity, for instance, near points at which $U$ diverges. Therefore, models in which $U(\xi)$ is singular in ${\xi=\xi_{sing}=\xi(|\R|\to\infty)}$ cannot be discussed within the framework of this paper. 

Also, it is clear from Eq.~\rif{eq:sol_alpha_explicit} that if $\dot\xi_0=\dot\xi_{\alpha,0}$ the amplitude of oscillations would vanish at all times. This can be immediately proved to be wrong, for instance numerically. This is an unfortunate consequence of the approximations used to derive \rif{eq:sol_alpha_explicit}, particularly neglecting $\ddot\xi_a$ in \rif{eq:ddot_xi_1}; evidently, the source term $\ddot\xi_a\neq 0$ will produce oscillations regardless of the initial conditions. When $\alpha$ is initially very small, $\ddot\xi_a$ and in general terms proportional to $1/t_{contr}^2$ should be kept and the approximations used are no longer valid. Therefore, Eq.~\rif{eq:sol_alpha_explicit} is reliable when $(\dot\xi_0-\dot\xi_{\alpha,0})$ is ``large'' enough, say of the order of $\dot\xi_{\alpha,0}$.

In order to have simple and more or less reliable estimates, we will use the initial conditions
\be\label{eq:initial_conditions}
\begin{cases}
 \R_0=-\tilde\T_0 \\
 \dot \R_0 = -\kappa\,\dot{\tilde\T_0} = -\kappa\,\T_0/t_{contr}\,,
\end{cases}
\ee
where $\kappa$ is a free parameter quantifying the initial displacement from the GR behaviour $\R+\tilde\T=0$; in particular, $\kappa=1$ corresponds to the situation in which $\R$ initially behaves exactly as if there were no $F(\R)$ at all (but still a cosmological constant). For simplicity, only change the initial ``velocity'' $\dot \R_0$.\\
Because of the considerations made above and noting that with these initial conditions
\be
\dot\xi_0 = \kappa\,\dot\xi_{a,0}\,,
\ee
our results will be particularly reliable for values of $\kappa$ not too close to unity.

\begin{figure}[h!]
\centering
 \includegraphics[width=\larghezza]{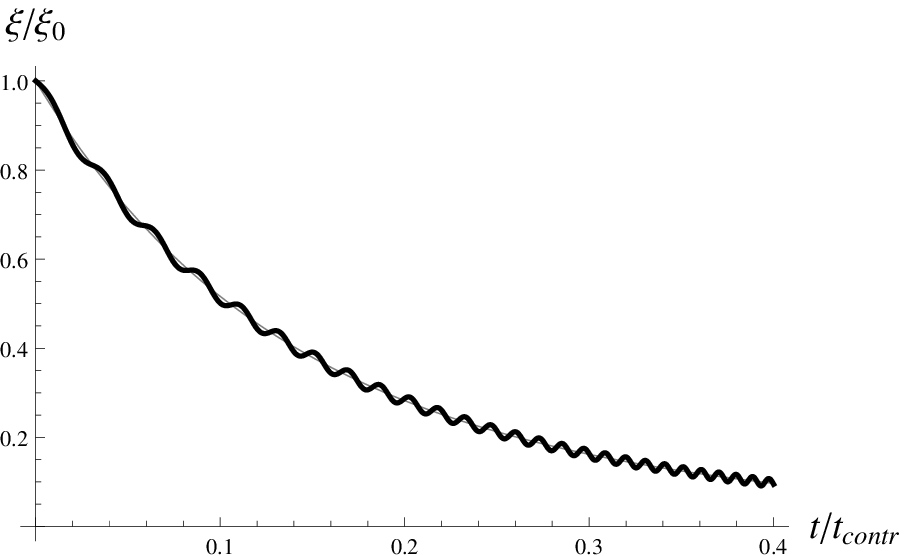}\\
\includegraphics[width=\larghezza]{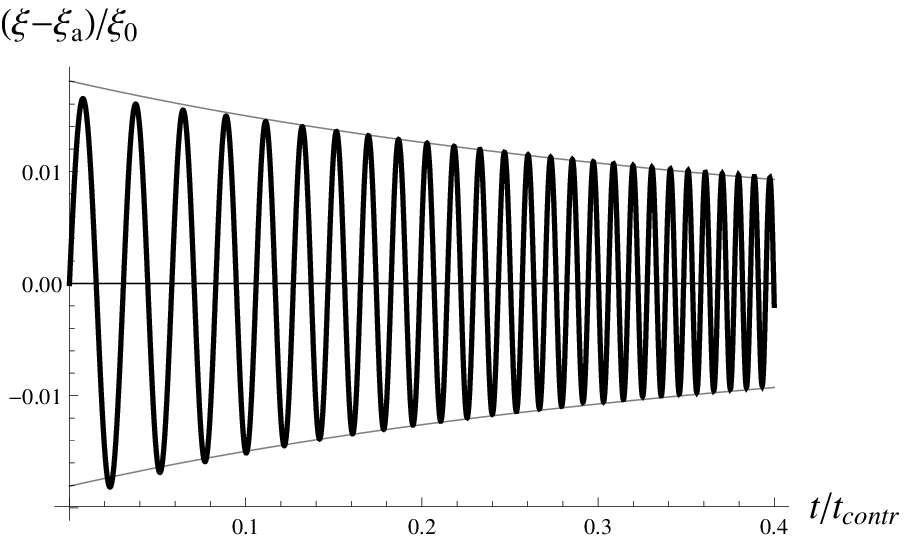}\\
 \includegraphics[width=\larghezza]{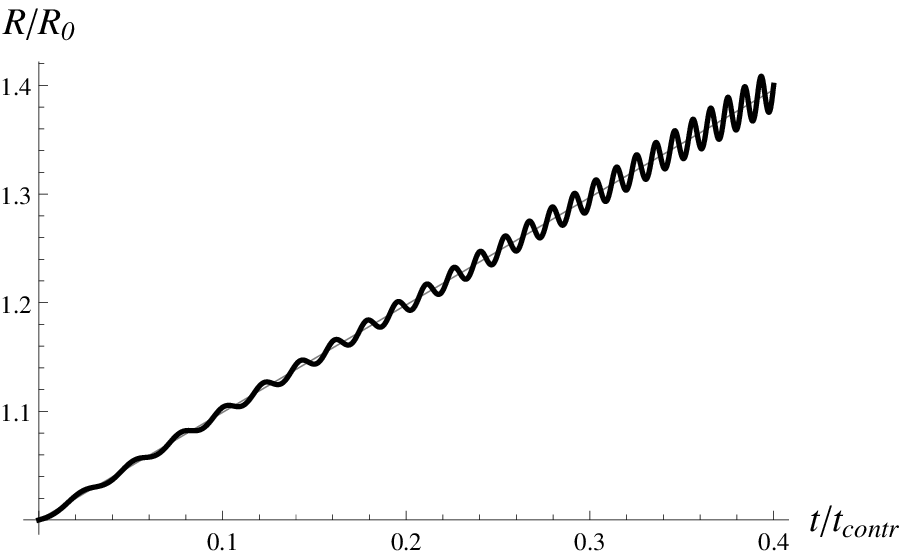}\\
\includegraphics[width=\larghezza]{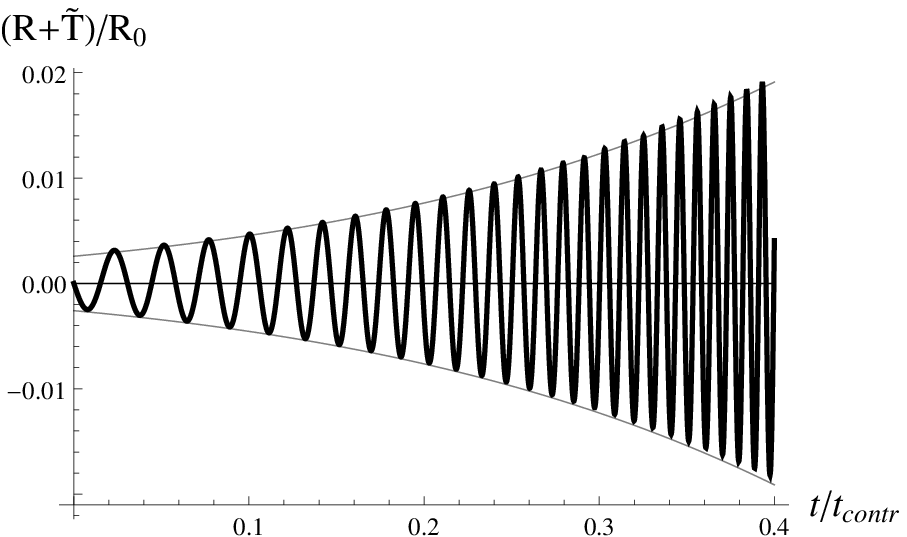}
\caption{Comparison of the evolution of $\xi$ for the Hu-Sawicki model with the predicted result, see equations~\rif{eq:xi_expansion} and \rif{eq:sol_alpha_explicit}. The values of parameters used are: ${n=3}$, ${\lambda =\R_{29}=1}$, ${\rho_{29}=2\cdot10^2}$ and ${t_{10}=1\cdot10^{-6}}$, ${\kappa=0.5}$.
The value $n=3$ gives satisfactory result for these models in reproducing the known cosmological evolution.
\i{Panel 1 (top):} numerical solution for $\xi$ (black), compared to the ``average'' value $\xi_a$ (gray), defined in~\rif{eq:xi_a_def}, normalised in units of $\xi_0$. \i{Panel 2:} plot of $\xi_1$~\rif{eq:xi_expansion}, compared to the expected evolution of the amplitude $\alpha$ \rif{eq:alpha_models_explicit}, normalised to $\xi_0$. \i{Panel 3:} evolution of $\R/\R_0$ with time (black), compared to the external energy/mass density (gray). Note that~\rif{eq:T_evol} is assumed. \i{Panel 4 (bottom)}: oscillations of $\R$ around its GR value $\R=-\tilde\T$; the amplitude grows with time following~\rif{eq:beta_fast_roll} quite accurately.}
\label{fig:xi_evol}
\end{figure}

\begin{figure}[h]
\centering
 \includegraphics[width=\larghezza]{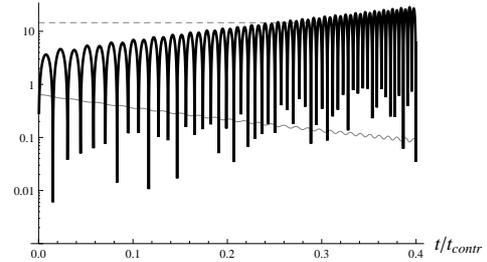}
\caption{Comparison of the different terms in~\rif{eq:trace}; the lines are, respectively, $|\R+\T|$ (black), $|2F|$ (dashed) and $10^{12}|F'\R|$ (gray), and parameters are those of Fig.~\ref{fig:xi_evol}. Notice that, as expected, $F'\R$ is absolutely negligible, whereas $F$ is of the order of $\R+\T$. Still, since it only produces an almost constant offset in the GR solution, there is no appreciable effect on solutions as indicated by fig.~\ref{fig:xi_evol}. For details, see the discussion above and below Eq.~\ref{eq:terms_com}.}
\label{fig:terms_comparison}
\end{figure}

Using these initial conditions, the amplitude of the scalaron oscillations for models~\rif{eq:models} evolves as (see also Eq.~\ref{eq:omega_models}):
\begin{subequations}\label{eq:alpha_models_explicit}
\begin{align}
\alpha(t) &\simeq \frac{\left[6n\lambda(2n+1)\right]^\frac{3}{2}|1-\kappa||\R_c|^{3n+\frac{3}{2}}\,\dot{\tilde\T_0}}{\tilde \T_0^\frac{5(n+1)}{2}\,\tilde\T(t)^{(n+1)/2}}\,,\\
& = \frac{\left[6n\lambda(2n+1)\right]^\frac{3}{2}|1-\kappa||\R_c|^{3n+\frac{3}{2}}\,T_0^2}{\tilde \T_0^\frac{5(n+1)}{2}\,\tilde\T(t)^\frac{n+1}{2}\,t_{contr}^2}\,.
\end{align}
\end{subequations}
Accordingly, $\R$ oscillates around its GR value $\R=-\tilde\T$. We thus define
\be\label{eq:R_expansion}
\R(t) = -\tilde\T + \beta\,\R_0\,r_{osc}\,,
\ee
where $r_{osc}$ has maximum absolute value equal to 1 and contains all the information about the oscillations of curvature, whereas the dimensionless function $\beta$ contains the information about the amplitude of such oscillations. Using~\rif{eq:xi_models_explicit} and~\rif{eq:alpha_models_explicit}, and expanding at linear order in $\beta$, we find
\be\label{eq:beta_fast_roll}
|\beta| \simeq \frac{\tilde\T^{2n+2}\,|\alpha|}{6n\lambda(2n+1)\,\tilde\T_0\,|\R_c|^{2n+1}}\,.
\ee
In figure \ref{fig:xi_evol} we show a comparison between the numerical solutions of~\rif{eq:trace_xi_KG} and our estimates, using the approximate model~\rif{eq:model_approx}. With the chosen values of parameters, we have
\be\label{eq:omega_t_contr}
\frac{\omega_0\,t_{contr}}{2\pi}\simeq 30\,,
\ee
so the fast-roll condition~\rif{eq:fast-roll-condition} is satisfied. The agreement between our analytical estimates and numerical results is expected to improve as ${\omega_0\,t_{contr}}$ increases. In Fig.~\ref{fig:terms_comparison} we show a comparison of the various terms in~\rif{eq:trace}.

\subsection{Approaching the Singularity: Anharmonic Oscillations}
As $\xi$ decreases and $|\R|$ increases, $\alpha$ grows so that eventually there appear anharmonic features in the oscillations of the scalaron. This is due to the fact that, when $\alpha$ becomes of the order of $\xi_a$, the higher-order terms in the potential, which had been neglected in~\rif{eq:potential_approx_harmonic}, become important. The different shape of the potential~\rif{eq:potentials_approx} on the left and on the right of the bottom (see also figure~\ref{fig:potentials}) determines an asymmetry of oscillations around the expected average value $\xi_a$. In particular, it is easy to infer that, redefining
\be
\xi_{min}\equiv \xi_a-\alpha_-\,,\qquad \xi_{max}\equiv \xi_a+\alpha_+\,,
\ee
we should have ${\alpha_-<\alpha_+}$, because the potential is steeper for ${\xi<\xi_a}$ than it is for ${\xi>\xi_a}$. Note that in the harmonic regime we assumed (see Eq.~\ref{eq:xi_a_+-_alpha})
\be
\xi_{max}-\xi_a = \xi_a-\xi_{min} =\alpha\,.
\ee
The variation of $\alpha$ is caused by the change in the shape of the potential with time and the increasing ``energy'' of the field, in the sense of equation~\rif{eq:energy_conservation}. In the harmonic region, using~\rif{eq:xi_min_xi_max} and~\rif{eq:potential_approx_harmonic} yields
\be
U(\xi_{max})\simeq U(\xi_{min}) \simeq U_0+\frac{1}{2}\,\omega^2\alpha^2\equiv U_0+\Delta U\,.
\ee
Note that all quantities involved are functions of time. The term $\Delta U$, if we neglect the integral term in~\rif{eq:energy_conservation}, corresponds to the maximum value of ${\dot\xi^2/2}$, that is the value this term has  when the field is at the bottom of the potential. Since we are basically considering a classical harmonic oscillator, this is an expected result. Substituting the explicit values, we find
\be\label{eq:delta_U}
\Delta U \simeq \frac{18\left[n\lambda(2n+1)(1-\kappa)\right]^2|\R_c|^{4n+2}\,\tilde\T^{n+1}\,\dot{\tilde\T_0}^2}{\tilde\T_0^{5n+5}}\,.
\ee
As mentioned before, this result depends essentially on the variation of the shape of the potential and on the increase of the energy of $\xi$, not on the assumption of harmonicity. Therefore, we will assume that $\Delta U$ continues to follow~\rif{eq:delta_U} even \i{away} from the harmonic region. In particular, we are interested in the region very close to the singularity, namely $\xi_a\simeq \alpha$. We will see numerically that this assumption is in good agreement with exact results.

Near the singularity, the term in the potential~\rif{eq:potentials_approx} linear in $\xi$ goes to zero more rapidly than the other term, so it can be neglected; therefore, the request that
\be
U(\xi_a-\alpha_-) = U_0 + \Delta U\,,
\ee
using equation~\rif{eq:xi_min_xi_max}, leads to the solution
\be\label{eq:alpha_singul_explicit}
\alpha_-(t) \simeq \xi_a(t)-6n\lambda\left[\frac{U_0(t)+\Delta U(t)}{3\lambda \R_c(2n+1)}\right]^\frac{2n+1}{2n}\,.
\ee
The explicit forms of $U_0$ and $\Delta U$ for the models considered are given, respectively, by equations~\rif{eq:U0} and~\rif{eq:delta_U}.

\begin{figure}[t]
\centering
\includegraphics[width=\larghezza]{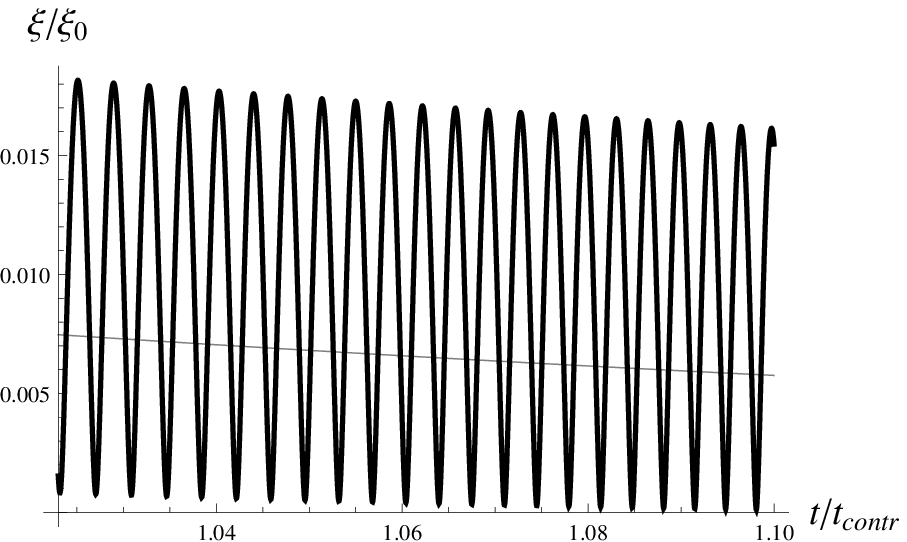}\\
\includegraphics[width=\larghezza]{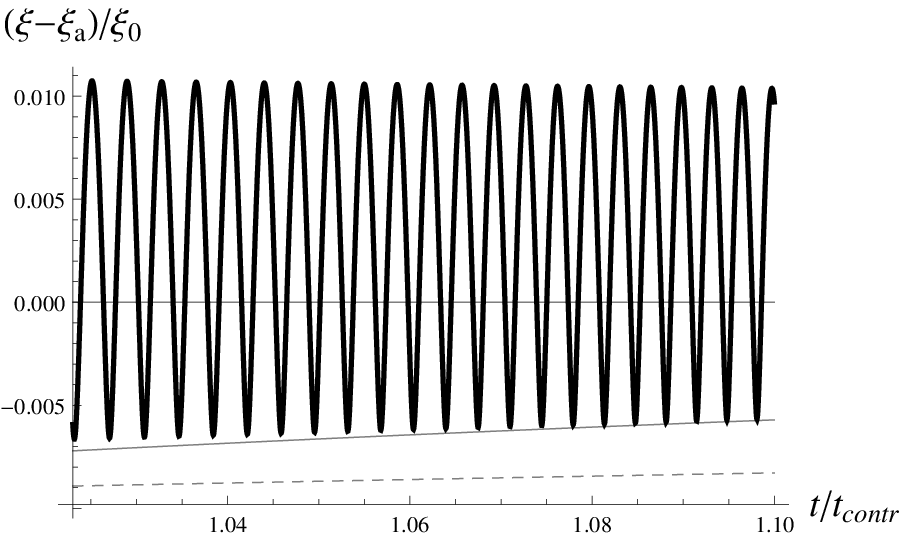}
\caption{Numerical solution for: $n=3$, $\lambda=\R_{29}=1$, $\rho_{29}=2\cdot 10^2$, $t_{10}=5\cdot10^{-7}$. \i{Top Panel:} as $\xi$ approaches the singularity, its oscillations around $\xi_a$ start showing anharmonic features.  \i{Bottom Panel:} the lower values of $\xi-\xi_a$ are clearly different from the naively predicted value~\rif{eq:alpha_models_explicit}, the dashed line in the picture, whereas the refined result of Eq.~\ref{eq:alpha_singul_explicit} (thin, solid line) is in very good agreement with the numerical solution.}
\label{fig:alpha_near_singularity}
\end{figure}

In figure~\ref{fig:alpha_near_singularity}, we show $\xi$ approaching the singularity, with ${\xi_{min}\ll \xi_a}$. As expected, the old estimate~\rif{eq:alpha_models_explicit} no longer reproduces the behaviour of numerical solutions, whereas the new result~\rif{eq:alpha_singul_explicit} works very well.

\subsection{Generation of the Singularity}\label{sec:fast_roll_singularity}
We are now ready to make the final calculations in order to derive the critical energy/mass density $\T_{sing}$ corresponding to the curvature singularity. We can either use~\rif{eq:alpha_singul_explicit} or equivalently the condition
\[
U_0+\Delta U = U(\xi_{sing}) = 0,
\]
to obtain
\begin{subequations}\label{eq:sing_fast_correct}
\be\label{eq:T_sing_fast_roll}
\begin{aligned}
\frac{\tilde\T_{sing}}{\tilde\T_0} &= \left[\frac{\tilde\T_0^{2n+4}}{6\lambda n^2(2n+1)^2(1-\kappa)^2\,|\R_c|^{2n+1}\,\dot{\tilde\T_0}^2}\right]^\frac{1}{3n+1}\\
&\simeq \left[0.28\,\frac{\rho_{29}^{2n+2}\,t_{10}^2\,(1+2\lambda \R_{29}/\rho_{29})^{2n+4}}{\lambda n^2(2n+1)^2(1-\kappa)^2\,\R_{29}^{2n+1}}\right]^\frac{1}{3n+1}\,.
\end{aligned}
\ee
The corresponding timescale for the formation of the singularity, using equation~\rif{eq:T_evol} and~\rif{eq:T_correction}, is simply
\be\label{eq:t_sing_fast_roll}
\frac{t_{sing}}{t_{contr}} = \frac{\T_{sing}}{\T_0}-1 = \frac{\tilde\T_{sing}-4\Lambda}{\tilde\T_0-4\Lambda}-1\,.
\ee
\end{subequations}
In table~\ref{tab:T_sing_fast_roll}, we show a comparison of a few analytical estimates with exact numerical results. The agreement increases with increasing $t_{contr}$ (and increasing $\T_{sing}$), which in fact corresponds to the situation in which the assumptions of adiabaticity are particularly reliable, see for instance Eq.~\ref{eq:fast-roll-condition}.\\
Even for the smallest values of $t_{contr}$ considered in Tab.~\ref{tab:T_sing_fast_roll}, the discrepancy between the numerical and analytical values is at most a few percent. Notice that this is a considerable and perhaps surprising result, since for the first value in table~\ref{tab:T_sing_fast_roll} ($t_{10}=1\cdot 10^{-6}$) we have
\[
\frac{\omega_0\,t_{contr}}{2\pi}\simeq 2\,,
\]
so that the condition~\rif{eq:fast-roll-condition} is actually barely fulfilled, and yet the analytical estimates work more than sufficiently well.

\begin{table*}[bt]
\begin{tabular}{c d d c c d d}
\toprule\noalign{\smallskip}
 &\multicolumn{2}{c}{$\qquad\tilde\T_{sing}/\tilde\T_0$} &  &  &\multicolumn{2}{c}{$\qquad\tilde\T_{sing}/\tilde\T_0$} \\
\noalign{\smallskip}
\cline{2-3}\cline{6-7}
\noalign{\smallskip}
$t_{10}$ & \multicolumn{1}{c}{$\qquad$Eq.~\rif{eq:sing_fast_correct}} & \multicolumn{1}{c}{$\qquad$exact} & $\qquad$ & $t_{10}$ & \multicolumn{1}{c}{$\qquad$Eq.~\rif{eq:sing_fast_correct}} &  \multicolumn{1}{c}{$\qquad$exact}\\
\noalign{\smallskip}
\colrule
\noalign{\smallskip}
$1\cdot 10^{-6}$ & 1.40877 & 1.56622 & & $1.0\cdot 10^{-5}$ & 2.23276  & 2.23157 \\
$2\cdot 10^{-6}$ & 1.61826 & 1.66162 & & $1.2\cdot 10^{-5}$ & 2.31567  & 2.3125 \\
$3\cdot 10^{-6}$ & 1.75495 & 1.77693 & & $1.4\cdot 10^{-5}$ & 2.38818  & 2.38475 \\
$4\cdot 10^{-6}$ & 1.85889 & 1.87618 & & $1.6\cdot 10^{-5}$ & 2.45282  & 2.44829 \\
$5\cdot 10^{-6}$ & 1.94373 & 1.95447 & & $1.8\cdot 10^{-5}$ & 2.51128  & 2.5066 \\
$6\cdot 10^{-6}$ & 2.01591 & 2.02037 & & $2.0\cdot 10^{-5}$ & 2.56476  & 2.55936 \\
$7\cdot 10^{-6}$ & 2.07903 & 2.08139 & & $2.5\cdot 10^{-5}$ & 2.68182  & 2.67566 \\
$8\cdot 10^{-6}$ & 2.1353  & 2.13753 & & $3.0\cdot 10^{-5}$ & 2.78141  & 2.77482 \\
$9\cdot 10^{-6}$ & 2.1862  & 2.18478 & & $5.0\cdot 10^{-5}$ & 3.0806   & 3.07293 \\

\noalign{\smallskip}\botrule
\end{tabular}
\caption{Critical energy/mass density $\T_{sing}$ obtained using~\rif{eq:T_sing_fast_roll}, compared to the exact numerical result. Parameters are $n=3$, $\lambda=\R_{29}=1$, $\kappa=0.5$, $\rho_{29}=10^2$, so results depend on the value of $t_{contr}$.}
\label{tab:T_sing_fast_roll}
\end{table*}

\begin{figure}[b]
\includegraphics[width=\larghezza]{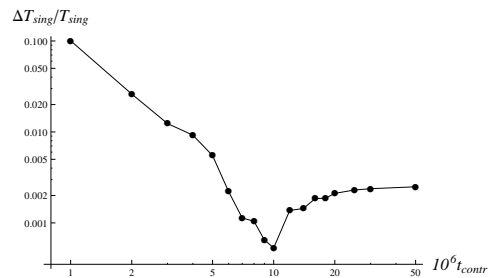}
\caption{Relative errors of table~\ref{tab:T_sing_fast_roll}. The discrepancy between analytical estimates and exact values decreases as $t_{contr}$ increases, but tends to a constant value of about 0.2\%. The ``cusp'' at small errors at $t_{10}\simeq 10^{-5}$ corresponds to a change in sign of $\Delta \T_{sing}/\T_{sing}$. See the text for further details.}
\label{fig:sing_error_fast_roll}
\end{figure}

In general, the accuracy of our analytical estimate should increase with increasing $\omega_0\,t_{contr}$, as the adiabatic approximation is more and more accurate. Instead, the relative errors seem to tend asymptotically (mind the logarithmic scale) to a fixed, though quite small, value ${~0.2\%}$, for which we have not found an explanation; possibly, this feature could be due to numerical computation issues. Anyway, the accuracy of our analytical estimates is good enough for all practical purposes.

One could argue that we should have displayed results for longer $t_{contr}$ and perhaps larger $\rho_0$, which are physically more realistic. Unfortunately, exploring that range of parameters is almost prohibitive from a computational standpoint, due to the huge number of oscillations occurring until $t$ reaches $t_{sing}$. Using equations~\rif{eq:omega_models} and~\rif{eq:sing_fast_correct} and assuming for simplicity $t_{sing}/t_{contr}>1$, we obtain in fact
\be
N_{osc} \simeq \int_{t_0}^{t_{sing}}\omega\,dt \propto \left(\rho_{29}^{n+1}\,t_{10}\right)^\frac{5n+5}{3n+1}\,,
\ee
so even a small increase in $\rho_{29}$ and/or $t_{10}$, especially for large $n$, can lead to an enormous increase in the time required for computations.

Nevertheless, we have no reason to believe that the satisfactory agreement of our estimates and numerical results would not hold in the case of more realistic values of parameters.

\section{Slow-Roll Region}\label{sec:xi_stuck}
Let us relax the assumptions of adiabaticity of section \ref{sec:adiabatic}, and focus instead on the opposite regime, that is
\be\label{eq:slow_roll_condition}
\frac{\omega\,t_{contr}}{2\pi}\ll 1\,,
\ee
corresponding to Eq.~\rif{eq:fast_roll_inequality} with inverted inequality sign. This is a slow-roll regime, in which the initial ``velocity'' of the field dominates over the acceleration due to the potential. In first approximation, assuming that $\dot\xi_0\neq 0$, which is equivalent to $\kappa\neq 0$ in~\rif{eq:initial_conditions}, we have
\be\label{eq:xi_slowroll_linear}
\xi(t) \simeq \xi_0+\dot\xi_0\,t\,.
\ee
Notice that $\dot\xi_0 < 0$. This behaviour, i.e. the fact that $\xi$ is roughly linear in $t$, can also be understood as follows. Considering equation~\rif{eq:trace_xi_KG}, we see that neglecting ${\R+\tilde\T}$ we are left with
\be\label{eq:trace_xi_slowroll}
\ddot\xi = 0\,,
\ee
which has exactly the solution~\rif{eq:xi_slowroll_linear}. This does not mean that we are precisely sitting on the solution $\R+\tilde\T=0$, because $\ddot\xi(\R+\tilde\T=0)\equiv\ddot\xi_a\neq 0$; otherwise, we would not have any singularity since $\R$ would simply follow the smooth evolution of $\tilde\T$. Rather, it means that~\rif{eq:trace_xi_slowroll} coincides with~\rif{eq:trace_xi_KG} up to corrections of order $\R+\tilde\T$. Since we can estimate
\be
\ddot\xi \sim \frac{\xi}{t_{contr}^2}\,,
\ee
and
\be
\R+\tilde\T = \frac{\partial U}{\partial\xi} \sim \omega^2\xi\,,
\ee
we find that
\be
\frac{\ddot\xi}{\R+\tilde\T} \sim \frac{1}{\omega^2t_{contr}^2} \gg 1\,.
\ee
This means that~\rif{eq:trace_xi_KG} and~\rif{eq:trace_xi_slowroll}, in this regime, are equal provided that we neglect terms of order $(\omega\,t_{contr})^2$; this is a legitimate approximation when~\rif{eq:slow_roll_condition} holds.

The reader may compare this to the assumptions of Sec.~\ref{sec:adiabatic}, where instead we had neglected terms ${\propto t_{contr}^{-2}}$. In that regime, the dominant contribution to $\ddot\xi$ was oscillatory, with
\[
\ddot\xi_{adiab} \sim \omega^2\xi\,,
\]
because we had $\omega\,t_{contr}\gg 1$.

\subsection{Generation of the Singularity}
\begin{figure}[t!]
\includegraphics[width=\larghezza]{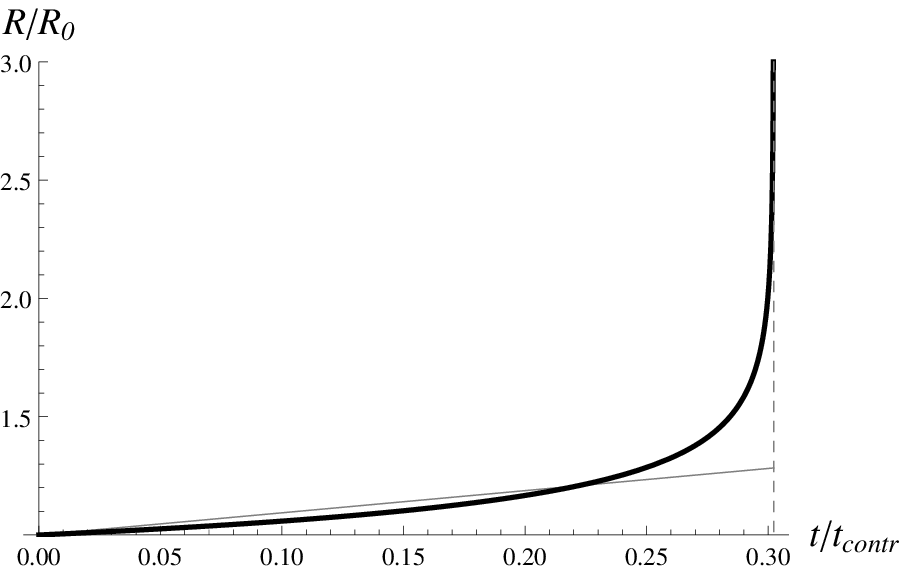}\\
\includegraphics[width=\larghezza]{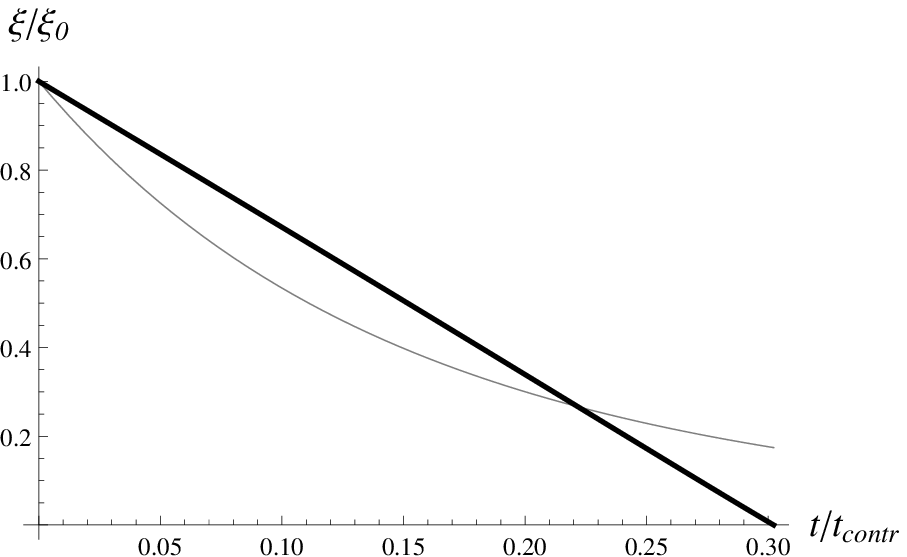}
\caption{Numerical results for parameters: ${n=3}$, ${\lambda=\R_{29}=1}$, ${\rho_{29}=30}$, ${t_{10}=1\cdot 10^{-5}}$, ${\kappa=0.5}$. Initially, ${\omega_0\,t_{contr}/2\pi\simeq 0.2}$, so the condition~\rif{eq:slow_roll_condition} is narrowly fulfilled. \i{Top Panel:} only a portion of $\R/\R_0$ (which diverges) is shown, in order to have a comparison with $\tilde\T/\tilde\T_0$ (thin, solid line); the singularity is reached at ${t_{sing}/t_{contr}\simeq 0.3}$ (dashed vertical line; see Tab.~\ref{tab:T_sing_slow_roll}). Initially, the slope of $\R/\R_0$ is different from that of $\tilde\T/\tilde\T_0$ because $\kappa\neq 1$. \i{Bottom panel}: $\xi$ does not follow $\xi_a$ (gray line) at all, but rather decreases roughly linearly with time, in qualitative agreement with~\rif{eq:xi_slowroll_linear}.}
\label{fig:singularity}
\end{figure}

With the simple solution~\rif{eq:xi_slowroll_linear}, it is straightforward to see that $\xi$ reaches the singularity $\xi_{sing}=0$ at
\be
t_{sing} \simeq -\frac{\xi_0}{\dot\xi_0}\,,\quad \tilde\T_{sing}\simeq \T_0\left(1-\frac{\xi_0}{\dot\xi_0\,t_{contr}}\right)+4\Lambda\,.
\ee
Using the explicit expressions for the models under consideration, we obtain the very simple expression
\begin{subequations}\label{eq:sing_slow_correct}
\be\label{eq:t_sing_slow_roll}
\frac{t_{sing}}{t_{contr}} \simeq \frac{1+4\Lambda/\T_0}{(2n+1)\kappa}\,,
\ee
or equivalently
\be\label{eq:T_sing_slow_roll}
\frac{\tilde\T_{sing}}{\tilde\T_0} \simeq 1 + \frac{1}{(2n+1)\kappa}\,.
\ee
\end{subequations}
In figure~\ref{fig:singularity}, we show the typical behaviour of $\xi$ and $R$ in this regime, until the singularity.
\hyphenation{initially}

\begin{figure}[b!]
\includegraphics[width=\larghezza]{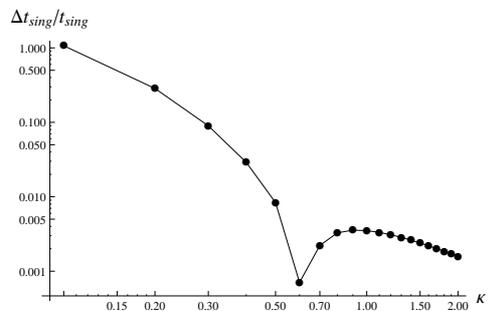}
\caption{Relative errors of table~\ref{tab:T_sing_slow_roll}. Noticeably, only for very small values of $\kappa$, say less than a few percent, there is an appreciable difference between our estimate and the exact result.}
\label{fig:sing_error_slow_roll}
\end{figure}

Basically, we are assuming that the motion of $\xi$ is completely dominated by the initial conditions, and that the acceleration due to the potential is negligible. Of course, for $\kappa\to 0$ the approximation loses its validity because the initial velocity is practically zero, but this is not worrisome because the most physically sensible choices are those with $\kappa\simeq 1$. The theoretical estimates of Tab.~\ref{tab:T_sing_slow_roll} are in remarkable agreement with the exact numerical values, and as expected the two results differ significantly only when $\kappa\ll 1$. The relative errors are depicted in Fig.~\ref{fig:sing_error_slow_roll}.

As expected, errors decrease for increasing values of $\kappa$, except for the region $\kappa\simeq 0.6$, which is most likely a numerical feature and should have no physical meaning. Nonetheless, the agreement between analytical and numerical values is excellent, especially considering that the slow-roll condition~\rif{eq:slow_roll_condition} is barely fulfilled, in fact ${\omega_0\,t_{contr}/2\pi\simeq 0.2}$.\\
The particular choice of parameters was motivated by the requirement of a somewhat realistic value of $t_{contr}$, in particular not too small. Taking larger values of $\rho_{29}$ and tuning $t_{10}$ to have, say, ${\omega_0\,t_{contr}/2\pi<1\%}$ yields estimates in outstanding agreement with numerical calculations, because the approximations~\rif{eq:large_density_limit} and~\rif{eq:slow_roll_condition} are all the more accurate. As an example, consider:
\be
\begin{cases}
n=3\\
\lambda=\R_{29}=1\\
\rho_{29}=10^2\\
t_{10}=10^{-9}\\
\kappa=1
\end{cases}
\quad \so \quad \frac{\omega_0\,t_{contr}}{2\pi}\simeq 0.2\%\,,
\ee
which gives the terrific value
\be
\frac{\Delta t_{sing}}{t_{sing}}\simeq  3\cdot 10^{-7}\,.
\ee
The price to pay, however, is to have unnaturally small contraction times, for instance ${t_{contr}=10}$ years in this case, therefore further similar results were not explicitly shown. Still, it is good to notice that the mathematical accuracy of our estimates improves as expected.

\begin{table*}[bth]
\begin{tabular}{c d d c c d d}
\toprule\noalign{\smallskip}
 &\multicolumn{2}{c}{$\qquad\,\,\, t_{sing}/t_{contr}$} &  &  & \multicolumn{2}{c}{$\qquad\,\,\,t_{sing}/t_{contr}$}\\
\noalign{\smallskip}
\cline{2-3} \cline{6-7}
\noalign{\smallskip}
\multicolumn{1}{c}{$\kappa$} & \multicolumn{1}{c}{$\qquad$Eq.~\rif{eq:sing_slow_correct}} & \multicolumn{1}{c}{$\qquad$exact} &  $\qquad\qquad$ & \multicolumn{1}{c}{$\kappa$} & \multicolumn{1}{c}{$\qquad$Eq.~\rif{eq:sing_slow_correct}} & \multicolumn{1}{c}{$\qquad$exact}\\
\noalign{\smallskip}\colrule\noalign{\smallskip}
0.1  & 1.52381   & 0.73706   &  & 1.1  & 0.138528  & 0.13899   \\
0.2  & 0.761905  & 0.592958  &  & 1.2  & 0.126984  & 0.127378  \\
0.3  & 0.507937  & 0.466093  &  & 1.3  & 0.117216  & 0.117551  \\
0.4  & 0.380952  & 0.370184  &  & 1.4  & 0.108844  & 0.109129  \\
0.5  & 0.304762  & 0.302237  &  & 1.5  & 0.101587  & 0.101831  \\
0.6  & 0.253968  & 0.253789  &  & 1.6  & 0.0952381 & 0.0954476 \\
0.7  & 0.217687  & 0.218171  &  & 1.7  & 0.0896359 & 0.0898169 \\
0.8  & 0.190476  & 0.191103  &  & 1.8  & 0.0846561 & 0.0848133 \\
0.9  & 0.169312  & 0.169917  &  & 1.9  & 0.0802005 & 0.0803378 \\
1.0  & 0.152381  & 0.152918  &  & 2.0  & 0.0761905 & 0.076311  \\


\noalign{\smallskip}\botrule
\end{tabular}
\caption{The parameters used are: ${n=3}$, ${\lambda=\R_{29}=1}$, ${\rho_{29}=30}$, ${t_{10}=1\cdot 10^{-5}}$.}
\label{tab:T_sing_slow_roll}
\end{table*}

\section{Remarks on the Validity of the Approximations Used}\label{sec:approx}
\subsection{Low Gravity: $g_\mn\simeq \eta_\mn$}
At the beginning of section~\ref{sec:CURV_EVOL}, we made the substitution
\be
\D^2\to\square\to\partial_t^2\,,
\ee
assuming the homogeneity of the cloud and low-gravity. The latter approximation is usually quite reasonable for astronomical densities, except for compact stars. However, one may be argue that if $\R\to\infty$, even with relatively low $\rho$, we are no longer in low-gravity regime, and thus the approximation fails. In order to show that $g_\mn\simeq \eta_\mn$ and $\R\to\infty$ are compatible, let us assume the simple homogeneous, isotropic line element
\be
ds^2 = dt^2 - \left[1+\psi(t)\right]d\mb x^2\,,\qquad |\psi|\ll 1\,,
\ee
with $\psi(t)$ parametrising the deviation from Minkowski. In this case, the scalar curvature is
\be\label{eq:R_psi}
\R = -\frac{3\ddot\psi}{1+\psi} \simeq -3\ddot\psi\,,
\ee
so we have
\begin{align}
\psi(t_{sing}) &\simeq -\frac{1}{3}\int^{t_{sing}}_{t_0}dt\int^{t}_{t_0}dt'\,\R(t') \notag\\
&\simeq \frac{1}{3}\int_{t_0}^{t_{sing}}dt\int^{t}_{t_0}dt'\,\tilde\T(t') \notag\\
&\simeq \frac{\tilde\T_0\,t_{sing}^2}{6}+\frac{\T_0\,t_{sing}^3}{18\,t_{contr}}\,\notag\\
&\simeq 0.28\,\rho_{29}\,t_{10}^2\,x^2\left(1 +\frac{x}{3}\right)\,.\label{eq:psi_sing}
\end{align}
where we have expanded $\R$ as in Eq.~\rif{eq:R_expansion}, neglected $\Lambda$ and defined
\be\label{eq:X_def}
x = \frac{t_{sing}}{t_{contr}}\,.
\ee
Low-gravity corresponds, roughly, to having~\rif{eq:psi_sing} smaller than unity. Firstly, we focus on the case $x\lesssim 1$, and assume that we are in the fast-roll (adiabatic) regime. Since in this case we can use Eq.~\rif{eq:sing_fast_correct}, the condition $x\lesssim 1$ becomes
\be
\frac{0.28\,\rho_{29}\,t_{10}^2}{\lambda n^2(2n+1)^2(1-\kappa)^2}\left(\frac{\rho_{29}}{\R_{29}}\right)^{2n+1}\lesssim 2^{3n+1}\,,
\ee
so that ${\rho_{29}\,t_{10}^2 \lesssim 1}$ as well, since ${\rho_{29}\gg \R_{29}}$. Therefore, 
\be
\psi \sim \rho_{29}\,t_{10}^2\,x < 1\,.
\ee
If on the other hand ${x\gtrsim 1}$, the condition ${\psi\lesssim 1}$ yields roughly
\be
\rho_{29}\,t_{10}^2\,x^3\lesssim 10\,,
\ee
that is
\[
\rho_{29}^{9n+7} \lesssim 4\cdot 10^{3n+1}\frac{\left[\lambda n^2(2n+1)^2(1-\kappa)^2\R_{29}^{2n+1}\right]^3}{t_{10}^{6n+8}}\,.
\]
It is easy to check that for all explicit numerical results presented in the text, this condition is very well satisfied. One should also keep in mind that when $t_{sing}/t_{contr}\gg 1$ the behaviour $\rho\sim t$ and thus the results of this paper are in any case expected to be less reliable.

The discussion of the slow-roll regime is even more straightforward, since $x<1$ (see Eqs.~\ref{eq:slow_roll_condition} and~\ref{eq:t_sing_slow_roll}) and
\be
\rho_{29}\,t_{10}^2 \sim \omega_0\,t_{contr}\left(\frac{\R_{29}}{\rho_{29}}\right)^{2n+1} \ll 1\,,
\ee
so $\psi$ is safely smaller than unity.

\subsection{Negligible Pressure: $p\ll\rho$}
Let us now consider the assumption of pressureless dust. We can combine the trace equation~\rif{eq:trace_approx} with the time-time component of the modified Einstein equations~\rif{eq:field_equations} assuming the equation of state $p=w\rho$, obtaining:
\be\label{eq:sys_Einstein}
\begin{cases}
\ddot\xi + \R + 2F = \cfrac{8\pi}{\mpl^2}\,(\rho -3 p) = \cfrac{8\pi(1-3w)\rho}{\mpl^2}\,,\\
\left(1-\cfrac{\xi}{3}\right)\R_{tt}-\cfrac{f}{2} = \cfrac{8\pi\,\rho}{\mpl^2}\,.\\
\end{cases}
\ee
Then, as in GR, the space-space equation is automatically fulfilled\footnote{For simplicity we have also assumed isotropy, that is ${\R_{xx} = \R_{yy} = \R_{zz}}$ and ${p_x=p_y=p_z}$, but the result can be easily generalised.}, hence:
\be
\left(1-\frac{\xi}{3}\right)\R_{ii}+\frac{f}{2}+\frac{\ddot\xi}{3}=\frac{8\pi p}{\mpl^2} = \frac{8\pi w \rho}{\mpl^2}\,.
\ee
This is true for any equation of state $w$, including of course the non-relativistic case $w=0$, which corresponds to assuming
\be
\frac{p}{\rho}\ll 1\quad\so\quad \T^\mu_\mu \simeq \frac{8\pi}{\mpl^2}\,\rho\,.
\ee
The mathematical consistency of the Einstein equations is therefore guaranteed regardless of the assumed equation of state. Physically, we know from statistical mechanics that for non-relativistic particles
\be
\frac{p}{\rho} \sim \frac{v^2}{3\,c^2}\,,
\ee
where $v$ is the typical velocity of the dust particles and $c$ is the speed of light. Given the total mass of the cloud $M$ and its radius
\be
r \simeq \left(\frac{3M}{4\pi\rho}\right)^{1/3}\,,
\ee
the velocity of the particles at time $t$ should approximately be
\be
v \sim |\dot r| \simeq \frac{r_0}{3\,t_{contr}(1+t/t_{contr})^{1/3}}\,.
\ee
Ultimately, this yields
\be
\frac{p}{\rho} \sim 10^{-9}\,\frac{M_{11}^{2/3}}{t_{10}^2\,\rho_{29}^{2/3}(1+t/t_{contr})^{8/3}}\,,
\ee
where
\be
M_{11} \equiv \frac{M}{10^{11}\,M_\odot} \simeq \frac{M}{2\cdot 10^{44}\,M_{11}\text{ g}}\,.
\ee
In basically any conceivable astronomical situation, except for very massive, rarefied clouds with short, perhaps unnatural contraction times, this quantity is much smaller than one, so that $p$ is indeed negligible.

We should be completely honest and point out that for the smaller values of $t_{contr}$ shown in table~\ref{tab:T_sing_fast_roll}, taking $M_{11}\sim 1$ gives a ratio $p/\rho>1$, which seems to invalidate the initial assumptions. However, because of the considerations at the end of section~\ref{sec:fast_roll_singularity}, we can disregard these problems provided that we carefully choose physically realistic parameters. In other words, some of the values in table~\ref{tab:T_sing_fast_roll} are unlikely to describe existing physical systems, but are nonetheless a useful indication of the accuracy of our analytical estimates.

\section{Discussion and Conclusions}
The possibility of curvature singularities in DE $f(\R)$ gravity models has been confirmed and studied in a rather simple fashion. The trace of the modified Einstein equations has been rewritten, under the simplifying assumptions of homogeneity, isotropy and low-gravity, as an oscillator equation for the scalaron field $\xi$, which moves in a potential $U$ depending on the external energy/mass density and thus on time. In the two models considered~\cite{Hu_Sawicki_2007,Starobinsky_2007}, the potential is finite in the point corresponding to the curvature singularity $|R|\to\infty$, that is $\xi_{sing}=0$; the energy conservation equation associated with $\xi$ indicates that the development of the singularity can be triggered by an increase in the external energy/mass density.

The ratio between the typical contraction time and the inverse frequency of the scalaron determines two distinct regimes. In the adiabatic regime the oscillations of $\xi$ are very fast compared to relevant variations of $U$, and such oscillations are almost harmonic. Performing a linear analysis, we have estimated the scalaron amplitude and frequency analytically. The singularity is expected to be reached when the amplitude of the oscillations of $\xi$ exceeds the separation between the ``average'' value $\xi_a$, which corresponds at each instant to the position of the bottom of the potential, and the singular point $\xi_{sing}$.\\
In the slow-roll regime, the typical oscillation time of the field $\xi$ is much longer than the typical contraction time, which also determines the timescale for significant changes in the potential. Thus, $\xi$ is mainly driven by its initial conditions, and the slope of the potential is not enough to stop the field from reaching the singular point. This may occur on relatively short timescales.

In both regimes, our analytical estimates and numerical results are in remarkable agreement (see tables~\ref{tab:T_sing_fast_roll} and~\ref{tab:T_sing_slow_roll}, and figures~\ref{fig:sing_error_fast_roll} and~\ref{fig:sing_error_slow_roll}).

In principle, the results of this work could provide simple methods to constrain and possibly rule out models~\cite{Hu_Sawicki_2007,Starobinsky_2007}, and most likely the same technique could be applied to other models already proposed as well as to more sophisticated evolution laws different from~\rif{eq:T_evol}. The development of a curvature singularity could reveal unexpected consequences in a more detailed analysis of the models, and the mechanisms described herein may play a highly non-trivial r\^ole, for instance, for the study of Jeans-like instabilities and hydrodynamical stellar (non-) equilibrium~\cite{Babichev_Langlois_and_Capozziello_De_Laurentis_2012}. This goes beyond the scope of this paper, and could be subject of further research.

Two effects could on one hand hinder the development of singularities, and on the other hand provide additional methods to constrain models: ultraviolet gravity modifications and gravitational particle production.\\
Ultraviolet corrections to the gravitational action should start dominating at large $\R$, and set a limit to its growth; in turn, $\R$ would never reach the singularity (for details see e.g.~\cite{Arb_Dolgov,Appleby_Battye_Starobinsky_2010}). Recently, a few works have investigated even more general ultraviolet aspects of (modified) gravity; a fully non-perturbative approach seems to point towards the altogether absence of singularities in gravity~\cite{Biswas_2011_and_Modesto_2012}.

Gravitational particle production, as is well known, is universal whenever curvature oscillates, and could in principle be a detectable source of high energy cosmic rays~\cite{Arb_Dolg_Rev_preparation}. The back-reaction on curvature is a damping of its oscillations, so this damping may prevent $\R$ from reaching infinity as well. This is particularly important in the adiabatic regime, where there can be very many oscillations before $\xi$ reaches $\xi_{sing}$ and therefore a large amount of energy could be released into SM particles. The produced cosmic rays would carry model-dependent signatures which could provide us valuable information to improve the constraints on the known models and maybe even suggest new gravitational theories.

\acknowledgments
The author is grateful to Prof. A.D. Dolgov and to an anonymous referee for useful discussions and criticism.

\end{document}